\documentclass[%
 reprint,
nofootinbib,
 amsmath,amssymb,
 aps,
]{revtex4-1}

\usepackage{lipsum}
\usepackage{amsmath}
\usepackage{braket}
\usepackage{graphicx}% Include figure files
\usepackage{dcolumn}% Align table columns on decimal point
\usepackage{bm}% bold math
\usepackage{hyperref}% add hypertext capabilities
\begin{document}

\title{Holography of negative energy states}% Force line breaks with \\
%\thanks{A footnote to the article title}%

\author{Felipe Rosso}
\email{felipero@usc.edu}
\affiliation{%
Department of Physics and Astronomy, University of Southern California,\\ Los Angeles, California 90089-0484, USA
}%

\begin{abstract}
Quantum states with negative energy densities have been long known to exist in quantum field theories. We explore the structure of such states for holographic theories using quantum information theory tools and show how certain negative energy states are naturally captured by the thermodynamics of black holes with hyperbolic horizon at zero temperature, suggesting that they provide a dual description of those states. Our results give a satisfying field theory understanding of the distinct thermodynamics of such black holes.
\end{abstract}

\maketitle

%%%%%%%%%%%%%%%%%%%%%%%%%%%%%%%%
\section{Introduction}
\label{sec:intro}
%%%%%%%%%%%%%%%%%%%%%%%%%%%%%%%%

Classical energy conditions are local inequalities  involving the energy-momentum tensor $T_{\mu \nu}$ which constrains the allowed matter in a classical theory; e.g., the null energy condition is given by $T_{\mu \nu}u^\mu u^\nu\ge 0$ with $u^\mu$ any null vector. Inequalities such as this one were first proposed in General Relativity in order to neglect unphysical solutions to Einstein gravity equations. They allow us to exclude exotic geometries such as wormholes \cite{Morris}, time machines \cite{Ori}, and warp drives \cite{Alcubierre:1994tu,Natario:2001tk}, while they are a key ingredient for proving some strong results such as singularity theorems \cite{Penrose,Hawking,Haw+Pen} and topological censorship \cite{Censorship}, among others.

When introducing quantum fields, it has been long known that such classical constraints fail to be true \cite{Epstein:1965zza,Fewster:2012yh}, since there are states in the Hilbert space with negative energy densities. In fact, the energy density at any given point in space-time can be made arbitrarily negative by choosing a suitable quantum state \cite{Fewster2003EnergyII,Fewster:2004nj}. One is then led to consider weaker energy constraints such as the averaged and quantum null energy conditions \cite{Bousso:2015mna,Faulkner:2016mzt,Balakrishnan:2017bjg}. To get a better grasp of the origin and relevance of such quantum bounds, it is important to understand the structure of these negative energy states. In this paper, we focus on their holographic description (see Ref. \cite{Lee:2016wcn} for previous work). To do so, we use quantum information theory techniques which have been previously shown to be very useful in the study of negative energy \cite{Blanco:2013lea,Blanco:2017akw,Wall:2012uf,Wall:2009wi}.

In the following section, we start by defining the \textit{modular vacua} of any global state reduced to a space-time region, as the states with a minimum expectation value on the modular hamiltonian of the reduced system. Using relative entropy, we show their similarities to the global vacuum of the theory. In Sec. \ref{sec:2}, we consider the ground state of a conformal field theory (CFT) reduced to a ball and show that the modular vacua maximize the amount of negative energy inside the ball and provide a sharp energy bound. 

In Sec. \ref{sec:3}, we present our main result, and show that for holographic CFTs the negative energy excitations and degeneracy of the modular vacua are naturally captured by the thermodynamics of black holes with a hyperbolic horizon at zero temperature. The thermodynamics of such black holes has long been know to have some odd features of which the interpretation has been for the most part unclear \cite{Emparan:1999gf,Cai:2001dz,Nojiri:2001aj}. Our results give a natural understanding of such behavior and suggest that these black holes provide a holographic description of the modular vacua of this setup.

%%%%%%%%%%%%%%%%%%%%%%%%%%%%%%%%
\section{Modular vacua}
\label{sec:1}
%%%%%%%%%%%%%%%%%%%%%%%%%%%%%%%%

We start with a general discussion regarding reduced states in which the modular vacua naturally appear. Consider an arbitrary quantum field theory in $d$-dimensional space-time and a fixed global state described by the density operator $\rho$. For any smooth and spacelike region $A$, we can define the reduced state as
\begin{equation}\label{eq:26}
\rho_A\doteq 
  {\rm Tr}_{\mathcal{H}_{\bar{A}}}\left(\rho\right)=
  \frac{e^{-K_A}}{Z}\ ,
  \qquad 
  Z\doteq {\rm Tr}_{\mathcal{H}_A}\left(e^{-K_A}\right)\ ,
\end{equation}
where $K_A$ is the modular hamiltonian and $\bar{A}$ is the complementary region with Hilbert space $\mathcal{H}_{\bar{A}}$. The operator $\rho_A$ usually has a more complex structure than the global state $\rho$, but describes the same physics when calculating the expectation value of an observable in $A$. More precisely, it verifies 
$$\langle \mathcal{O}_A \rangle=
  {\rm Tr}_{\mathcal{H}_A \otimes \mathcal{H}_{\bar{A}}}\left(\rho \,\mathcal{O}_A\right)=
  {\rm Tr}_{\mathcal{H}_A}\left(\rho_A \mathcal{O}_A\right)\ ,$$
where $\mathcal{O}_A$ is any operator in the causal domain of $A$. By considering $\rho_A$ instead of $\rho$ we become independent of the degrees of freedom in $\bar{A}$ at the expense of considering a more complicated density operator. In this context, a natural question that arises is what is the pure state $\ket{\psi}$ we can construct in the Hilbert space $\mathcal{H}_A$ that is the most ``similar" to $\rho_A$. Relative entropy, defined as
\begin{equation}\label{eq:40}
S(\rho_1||\rho_2)\doteq 
  {\rm Tr}\left(\rho_1\ln(\rho_1)\right)-
  {\rm Tr}\left(\rho_1\ln(\rho_2)\right)
\end{equation}
for any density operators $\rho_1$ and $\rho_2$, seems to be particularly well suited to answer such a question since it is a measure of the statistical distance between $\rho_1$ and $\rho_2$ in the following sense: given the state $\rho_1$, the probability of  confounding  it with state $\rho_2$ after $N$ trials of some measurement decays as $e^{-NS(\rho_1||\rho_2)}$ for large $N$ \cite{Vedral}. It therefore allows for a precise quantification on how similar a state $\ket{\psi}$ is to $\rho_A$. 

We then consider (\ref{eq:40}) with $\rho_2=\rho_A$ and $\rho_1=\ket{\psi}\bra{\psi}$. Writing $\rho_A$ in terms of its modular hamiltonian $K_A$ and using that the entanglement entropy of $\rho_1$ vanishes since it describes a pure state, we find
\begin{equation}\label{eq:15}
S(\rho_1||\rho_A)=
  \bra{\psi} K_A \ket{\psi}-
  \langle K_A \rangle_{\rho}
  +
  S(\rho_A)
  \ .
\end{equation}
To calculate the first term we use that $K_A$ is a hermitian operator, meaning that it will be diagonalized by a complete and orthonormal set $\left\lbrace \ket{\psi_w(u)} \right\rbrace$ with real eigenvalues $k(u)$, where $u$ and $w$ are parameters which label the eigenspace and its degeneracy respectively. Expanding $\ket{\psi}$ in this set, 
$$\ket{\psi}=
  \int dwdu\,
  g(u,w)\ket{\psi_w(u)}\ ,
  \quad  
  \int dwdu\,|g(u,w)|^2=1\ ,$$
the relative entropy in (\ref{eq:15}) becomes
\begin{equation}\label{eq:7}
S(\rho_1||\rho_A)=
  \int dwdu\,
  k(u)|g(u,w)|^2
  -
  \langle K_A \rangle_{\rho}
  +
  S(\rho_A)
  \ .
\end{equation}
We can further simplify this expression by writing $k(u)$ in terms of the Renyi entropies of $\rho_A$, defined as
\begin{equation}\label{eq:39}
S_q(\rho_A)\doteq 
  \frac{1}{1-q}\ln
  \left[\frac{{\rm Tr}_{\mathcal{H}_A}\left(e^{-qK_A}\right)}{Z^q}\right]\ ,
\end{equation}
with $q\in \mathbb{N}_0$. The following values of $q$ are particularly useful,
$$
S(\rho_A)=
  \langle K_A \rangle_{\rho}+\ln(Z)\ ,
  \qquad 
  S_{\infty}(\rho_A)=
  k_0+\ln(Z)\ ,
$$
where $S_{q=1}(\rho_A)= S(\rho_A)$ is the entanglement entropy and $k_0 \doteq k(u_{\rm min})$ is the minimum eigenvalue of $K_A$, which can be written as
\begin{equation}\label{eq:36}
  k_0=
  -\big(S(\rho_A)
  -S_\infty(\rho_A)
  \big)+\langle K_A \rangle_{\rho}\ .
\end{equation}
Since (\ref{eq:7}) will be minimum when $k(u)=k_0$, we can use expression (\ref{eq:36}) and find
\begin{equation}\label{eq:43}
S(\rho_1^{\rm min}\,||\rho_A)=
  S_\infty(\rho_A)\ ,
  \qquad 
  \ket{\psi^{\rm min}}=\int dw\,g(w)\ket{\psi_w^{\rm min}}\ ,
\end{equation}
where $\ket{\psi_{w}^{\rm min}}$ are the eingestates of $K_A$ with minimum eigenvalue $k_0$. We conclude that any linear combination of $ \ket{\psi_w^{\rm min}}$ minimizes the statistical distance to $\rho_A$ over the set of pure states in $\mathcal{H}_A$. Just from the definition of the modular hamiltonian (\ref{eq:26}), this is a very natural result and is in accordance with the behavior of a thermal state $e^{-\beta H}/Z_\beta$, where the ground state $\ket{0}$ (which has the minimum eigenvalue of energy $H\ket{0}=0$) is also the closest pure state. 

This analogy is in fact quite precise as can be seen from defining the following unitary operator $U(s)=e^{isK_A}/Z$. Considering the action $\mathcal{O}_A\rightarrow \mathcal{O}_A(s)=U(s)\mathcal{O}_AU(-s)$ on any operator $\mathcal{O}_A$, we can formally prove that $\rho_A$ is thermal with respect to translations in $s$, by showing that it satisfies the Kubo-Martin-Schwinger (KMS) periodicity condition\footnote{The KMS periodicity condition provides a formal definition of a thermal state for operators in infinite-dimensional space. To show it holds with $\beta=1$, notice that $U(i)=\rho_A^{-1}$ and $U(-i)=\rho_A$.}
$$
{\rm Tr}_{\mathcal{H}_A}
  \Big(
  \rho_A\mathcal{O}_A(s+i\beta)\widetilde{\mathcal{O}}_A
  \Big)=
  {\rm Tr}_{\mathcal{H}_A}
  \Big(
  \rho_A\widetilde{\mathcal{O}}_A\mathcal{O}_A(s)
  \Big)\ ,
$$
for any operators $\mathcal{O}_A$ and $\widetilde{\mathcal{O}}_A$ and inverse temperature $\beta=1$. It is then reasonable not only to refer to the states $\ket{\psi_w^{\rm min}}$ in (\ref{eq:43}) as the \textit{modular vacua} of the reduced system but also to call the expectation value $\langle K_A \rangle$ the \textit{modular energy}. 

The modular vacuum energy is given by $k_0$ (\ref{eq:36}) and provides a sharp bound for the expectation value of $K_A$ on any state
\begin{equation}\label{eq:44}
\langle K_A \rangle \ge \bra{\psi^{\rm min}_w}K_A\ket{\psi^{\rm min}_w}=k_0\ . 
\end{equation}
Calculating $k_0$ explicitly for a particular system gives an inequality that can supply interesting information about the field theory under consideration. In the following, we will consider this inequality for a particular system and show that it gives a constraint on the negative energy excitations on the causal domain of $A$. 

%%%%%%%%%%%%%%%%%%%%%%%%%%%%%%%%
\section{Negative energy bound}
\label{sec:2}
%%%%%%%%%%%%%%%%%%%%%%%%%%%%%%%%

The previous discussion was done in full generality for any state $\rho$ and quantum field theory. To further investigate the structure of the modular vacua, we consider the global ground state $\rho=\ket{0}\bra{0}$ of a CFT in $d$-dimensional Minkowski space-time and take the region $A$ as a ball of radius $R$, so that the modular hamiltonian is given by \cite{Casini:2011kv,Bisognano:1976za}
\begin{equation}\label{eq:29}
K_A=
  \int_{\mathcal{C}_A}d\Sigma^\nu \,
  \xi^\mu
  T_{\mu \nu} \ ,
\end{equation}
where $d\Sigma^\nu=d\Sigma \,n^\nu$, with $n^\nu$ a unit vector normal to any $(d-1)$-dimensional spacelike surface $\mathcal{C}_A$ in the causal domain of the ball of which the boundary is at $t=0$ and $|\vec{x}|=R$. The conformal Killing vector $\xi^\nu$ generates a flow that keeps the sphere fixed and is given by 
\begin{equation}\label{eq:45}
\xi=
  2\pi
  \left(
  \frac{\left(R^2-|\vec{x}|^2-t^2\right)\partial_t-
  2tx^i\partial_i}{2R}
  \right)\ .
\end{equation}
It can be interpreted as an inverse local temperature vector, which can be defined and calculated for much more general systems \cite{Arias:2016nip,Arias:2017dda}. 

Considering different surfaces $\mathcal{C}_A$ will change the explicit expression of $K_A$ but leave its spectrum unchanged.\footnote{See Sec. 2.1 of Ref. \cite{Casini:2009vk}.} For definiteness, we may take $\mathcal{C}_A$ at $t=0$ so that the modular hamiltonian can be written as
\begin{equation}\label{eq:37}
K_A=2\pi
  \int_{|\vec{x}| \le R}
  d^{d-1}x\,
  \left(\frac{R^2-|\vec{x}|^2}{2R}\right)T_{00}(\vec{x})\ .
\end{equation}
This operator gives the energy density in the ball as weighted by the inverse local temperature, which is a positive function. Due to local negative energy excitations, we expect this operator to have some negative eigenvalues in its spectrum. The modular vacua correspond to a very special set of states, given by the ones which maximize the amount of negative energy in the ball. From (\ref{eq:36}) we already see that their modular energy $k_0$ will be negative, since the Renyi entropy is a decreasing function of $q$ and $\langle K_A \rangle_\rho=\bra{0} K_A \ket{0}=0$. Moreover, from (\ref{eq:44}), we have the following inequality
\begin{equation}\label{eq:38}
\int_{\mathcal{C}_A}d\Sigma^\nu \,
  \xi^\mu 
  \langle T_{\mu \nu} \rangle  
  \ge 
  k_0=
  -\big(S(\rho_A)
  -S_\infty(\rho_A)
  \big)\ ,
\end{equation}
which holds for the expectation value of any state  and surface $\mathcal{C}_A$, and the bound is sharp for the modular vacua. The modular vacuum energy $k_0$ gives a bound on the negative energy excitations in the causal domain of $A$. The fact that (\ref{eq:38}) holds for an infinite set of surfaces $\mathcal{C}_A$ is specially interesting.

The right-hand side of this inequality will not only be negative but also divergent, due to the infinite entanglement contributions captured by the Renyi entropies on the boundary of the ball. Just from the integral expression on the left-hand side such a behavior is not a surprise and can be expected. 

The key observation is the fact that, when considering averages of energy densities, the weight function should be defined in a complete Cauchy surface.\footnote{For example, in (\ref{eq:37}), it should be defined in the whole space.} Therefore, in order to recover the integral expression in (\ref{eq:38}), such a function must be equal to zero outside the ball and given by the inverse local temperature inside. Since the conformal Killing vector (\ref{eq:45}) vanishes at the boundary $|\vec{x}|=R$, the resulting weight function is continuous, but nondifferentiable. This apparently minor and technical detail is the reason the integral (\ref{eq:38}) is able to capture infinite negative energy excitations on the boundary of the ball and become divergent for certain quantum states. This was explicitly shown by Fewster and Hollands (Sec. 4.2.4 of Ref. \cite{Fewster:2004nj}) and Verch (Proposition 3.1 of Ref. \cite{Verch:1999nt}) for two-dimensional CFTs, and we will provide additional evidence in Appendix \ref{sec:apA}.\footnote{There is also evidence that, even for smooth weight functions in $d>2$, such integrals can be divergent because the average is over a spacelike surface. See Refs. \cite{Ford:2002kz,Fewster:2002ne} for explicit examples for scalar fields in $d=4$.} 

Apart from having an understanding of the divergence on both sides of (\ref{eq:38}), we learn that both have their origin in the sharp localization of boundary of the region. On this boundary, the Renyi entropy captures infinite entanglement contributions while the integral, infinite negative energy excitations. 

Despite this divergent behavior, the derived energy bound is still an interesting quantity to study, especially because it is sharp for the modular vacua. We will illustrate this in the following section by showing how nontrivial information can be extracted from it. There are other energy inequalities, such as the quantum null energy condition \cite{Balakrishnan:2017bjg,Bousso:2015mna}, which are useful and conceptually interesting despite of the fact that for certain states they involve divergent quantities \cite{Fu:2017ifb}. 

In Appendix \ref{sec:apA}, we use an independent approach to rederive, generalize, and calculate explicitly the inequality (\ref{eq:38}) for two-dimensional CFTs.

The modular vacua seem to be given by a complex set of states which are very difficult to study using standard field theory tools. In the following section, we will show that, when considering holographic CFTs, these states are captured in a very simple way by hyperbolic black holes at zero temperature.

%%%%%%%%%%%%%%%%%%%%%%%%%%%%%%%%
\section{Holography of the modular vacua}
\label{sec:3}
%%%%%%%%%%%%%%%%%%%%%%%%%%%%%%%%

We now explicitly compute the modular vacuum energy $k_0$ for this system. To do so, we use the construction developed in Ref. \cite{Casini:2011kv}, where it was shown that the reduced ground state on the ball $\rho_A$ can be conformally mapped to a thermal state with temperature $\tilde{T}=1/(2\pi R)$ on a background geometry $\mathbb{R}\times \mathbb{H}^{d-1}$, where $\mathbb{H}^{d-1}$ is a hyperbolic plane with curvature scale $R$. Given that $\rho_A$ and the thermal state are related by a unitary conformal transformation, the Renyi entropy (\ref{eq:39}) is invariant and can be calculated from the free energy of the thermal state as \cite{Hung:2011nu}
\begin{equation}\label{eq:21}
S_{q}(\rho_A)=
  -\left(
  \frac{F(\tilde{T}/q)-F(\tilde{T})}{\tilde{T}/q-\tilde{T}}
  \right)\ ,
\end{equation}
where $F(T)\doteq E(T)-TS(T)$, with $E(T)$ and $S(T)$ the energy and entropy of the thermal state. In particular, the entanglement entropy and infinite Renyi entropy are given by
\begin{equation}\label{eq:27}
S(\rho_A)=S(\tilde{T})\ ,
  \qquad 
  S_{\infty}(\rho_A)=S(\tilde{T})+
  \frac{E(0)-E(\tilde{T})}{\tilde{T}}
  \ .
\end{equation}
Using these expressions in (\ref{eq:38}), $k_0$ can be written as
\begin{equation}\label{eq:17}
  k_0=
  \frac{E(0)-E(\tilde{T})}{\tilde{T}}
\ .
\end{equation}
For an arbitrary CFT, this result is not particularly useful, since the calculation of the energy of a thermal state in a hyperbolic geometry is still a very difficult computation. However, if we restrict to holographic CFTs, the AdS/CFT dictionary \cite{Maldacena:1997re,Gubser:1998bc,Witten:1998qj} suggests that the thermal state will be dual to a black hole in asymptotic anti-de Sitter (AdS) with a hyperbolic horizon. This means that the energy of the thermal state is mapped to the mass of the black hole $E(T)\rightarrow M_{\rm BH}(T)$, a quantity that can be obtained from a standard computation. 

For a generic temperature, the mass of the hyperbolic black hole will depend on the gravity theory to which the specific CFT is dual. However, in Ref. \cite{Casini:2011kv}, it was shown that for $T=\tilde{T}$ the thermal state is described by a hyperbolic slicing of AdS, which has a finite temperature $\tilde{T}$ associated to an acceleration horizon analogous to Rindler's in Minkowski space-time. Since pure AdS is a solution to any covariant theory of gravity with negative cosmological constant, the above result is completely general. Moreover, the ``mass" of pure AdS vanishes $M_{\rm BH}(\tilde{T})=0$, meaning that the modular vacuum energy can be computed holographically as
$$
  k_0=M_{\rm BH}(0)/\tilde{T}\ ,
$$
where $M_{\rm BH}(0)$ is the zero temperature mass of the black hole solution with a hyperbolic horizon in the dual gravity theory.\footnote{For even dimensions, the black hole mass at $T=\tilde{T}$ might not be zero but have a constant Casimir contribution; see Ref. \cite{Emparan:1999pm}. This will have no impact in our discussion since $k_0$ is given by the difference between masses.} 

This expression might seem peculiar, given that in the previous section we argued that $k_0$ should not only be negative but divergent, which seems a curious thing to expect from the zero temperature mass of a black hole. However, it has long been known that black holes in asymptotic AdS with a hyperbolic horizon have an exceptional thermodynamics in which their zero temperature mass has exactly these characteristics: it is negative and divergent \cite{Emparan:1999gf,Cai:2001dz,Nojiri:2001aj}. The most negative value of mass allowed by the black hole thermodynamics is given by $M_{\rm BH}(0)$, in exact correspondence with the maximum amount of negative energy allowed by the theory inside the ball according to (\ref{eq:38}). We have therefore found a very satisfying holographic explanation for the unusual thermodynamics of hyperbolic black holes in asymptotic AdS.

We can also investigate how the degeneracy of the modular vacua $\Omega_0$ is encoded in the black hole thermodynamics. This was already considered in Sec. 5 of Ref. \cite{Hung:2011nu} by comparing the large $q$ expansion of the Renyi entropy expressions (\ref{eq:39}) and (\ref{eq:21}), where a simple calculation shows
$$
\ln(\Omega_0)=S_{\rm BH}(0)
  \ .
$$
This means that if we consider a flat superposition of the modular vacua
\begin{equation}\label{eq:28}
\rho_0=
  \int \frac{dw}{\Omega_0} \ket{\psi_w^{\rm min}}\bra{\psi_w^{\rm min}}\ ,
\end{equation}
we have
\begin{subequations} 
\label{eq:31}
\begin{eqnarray}
\langle K_A \rangle_{\rho_0}=
  {\rm Tr}_{\mathcal{H}_A}\left(\rho_0K_A\right)=
  M_{\rm BH}(0)/\tilde{T}\ ,
  \\ [5pt]
S(\rho_0)=-{\rm Tr}_{\mathcal{H}_A}\left(\rho_0\ln(\rho_0)\right)=S_{\rm BH}(0)\ .
\end{eqnarray}
\end{subequations}
We emphasize that these expressions hold for \textit{any} holographic CFT and therefore suggest the following: the hyperbolic black holes at zero temperature provide a holographic description of a flat superposition of the modular vacua of the ground state of a CFT reduced to a ball (\ref{eq:28}). This is in line with the field theory discussion of Sec. \ref{sec:1}, where we pointed out the similarities between the modular vacua and the ground state; both their holographic duals, pure AdS and the hyperbolic black hole, are at zero temperature.

For a specific gravity theory, the mass and entropy of the black hole can be computed and written in terms of field theory quantities through standard methods. In Appendix \ref{sec:apB}, we briefly review the calculation for Einstein gravity. The procedure is similar to the ones presented in Ref. \cite{Hung:2011nu}, and in fact, some results can already be extracted from their equations through (\ref{eq:38}).

By considering the hyperbolic black hole solution in Einstein gravity \cite{Emparan:1999gf} and using (\ref{eq:31}), we find 
\begin{subequations} 
\label{eq:20}
\begin{eqnarray}
\langle K_A^{(E)} \rangle_{\rho_0}=&
  \left(\frac{1-d}{d}\right)
  \left(\frac{d-2}{d}\right)^{(d-2)/2}S(\rho_A)\ ,
  \label{eq:12} \\[5pt]
S(\rho_0^{(E)})=&
  \left(\frac{d-2}{d}\right)^{(d-1)/2}
  S(\rho_A) \ , \label{eq:16}
\end{eqnarray}
\end{subequations}
where $S(\rho_A)$ is the entanglement entropy of $\rho_A$. As expected, the modular vacuum energy is negative and divergent since it is proportional to $S(\rho_A)$. The degeneracy is also divergent apart from the $d=2$ case where the modular vacuum is unique, in agreement with Ref. \cite{CalabreseLefevre}.

Since not all holographic field theories will be dual to Einstein gravity, we can also consider the Gauss-Bonnet hyperbolic black hole \cite{Cai:2001dz,Nojiri:2001aj} for $d\ge 4$, which allows for field theories with a more complicated structure. Although the mass and entropy can be computed analytically for generic $d$, the expressions are quite complicated, so we only present the $d=4$ results, which are given by
\begin{subequations} 
\label{eq:19}
\begin{eqnarray}
\langle K_A^{(GB)} \rangle_{\rho_0}
=&
  4n_c^2/(5n_c-1)\langle K_A^{(E)}\rangle_{\rho_0}
  \ ,
  \\[5pt]
S(\rho_0^{(GB)})
=&
\frac{(-3n_c^2+6n_c-1)\sqrt{8(3n_c-1)}}{(5n_c-1)^{3/2}}S(\rho_0^{(E)})
   , \label{eq:23}
\end{eqnarray}
\end{subequations}
where $n_c\doteq c/a$ with $a$ and $c$ the central charges in $d=4$, defined in the usual way from the trace of $\langle T_{\mu \nu} \rangle$. The allowed range of $n_c$ is given by $n_c\in [2/3,1+\sqrt{2/3}]$ (see Appendix \ref{sec:apB} for details). 

We can consider the behavior of these quantities for a fixed value of $a$ and variable $c$. Since the entanglement entropy is independent of $c$ \cite{Casini:2011kv}, from (\ref{eq:19}), we can directly analyze how the modular vacuum energy and degeneracy behave as a function of $c$. As $c$ increases, so does the modular vacuum energy, while its degeneracy decreases and becomes equal to 1 for $n_c=1+\sqrt{2/3}$. This behavior together with the energy inequality (\ref{eq:38}) means that, while CFTs with larger $c$ allow for more negative energy inside the ball, the number of states with this critical behavior decreases. This is a nontrivial statement that we were able to extract from the bound (\ref{eq:38}) despite its divergent nature.

%%%%%%%%%%%%%%%%%%%%%%%%%%%%%%%%
\section{Discussion}
%%%%%%%%%%%%%%%%%%%%%%%%%%%%%%%%

In this work, we have explored the holographic description of the modular vacua of the ground state of a CFT reduced to a ball, which contain maximum amount of negative energy inside this region. Despite the fact that such states seem to have a very complicated structure which makes them difficult to study using field theory techniques, we have shown through (\ref{eq:31}) that their holographic counterpart seems quite simple and given by hyperbolic black holes at zero temperature. The negative mass of such black holes played a crucial role in capturing the negative energy excitations.

Though our analysis was made entirely for zero temperature black holes, we can speculate on the holographic meaning of finite temperatures. Pure AdS (which has zero temperature) is dual to the ground state of the CFT, while thermal excitations are described by a black hole at finite temperature. Given the similarities between the ground state and the modular vacua discussed in Section \ref{sec:1}, we might consider an analogous situation; the modular vacua are dual to the zero temperature hyperbolic black hole, while excitations of those modular vacua are described by the finite temperature black hole. Since its mass will be negative for temperatures between zero and $\tilde{T}$ (where the mass vanishes $M_{\rm BH}(\tilde{T})= 0$), such a range could correspond to other states in the CFT with negative energy inside the ball. For small perturbations of the $\tilde{T}$ case toward smaller temperatures, a simple argument suggests that this is indeed so (see Sec. 4.2 of Ref. \cite{Johnson:2018amj}).

A crucial step for making the connection at zero temperature was the large $q$ expansion of the Renyi entropy. A further analysis of the subleading contributions of the expansions obtained from its usual definition (\ref{eq:39}) and the thermodynamic expression (\ref{eq:21}) might shed some light onto the meaning of hyperbolic black holes at small but finite temperature. 

From the field theory perspective, it is also interesting to continue the study of the modular vacua for systems in which the modular hamiltonian has nonlocal contributions. Though it is unclear whether such states will still have negative energy density inside the region, inequality (\ref{eq:44}) might contain interesting physical information. A good starting point for this analysis is to consider a two-dimensional free chiral fermion or scalar field reduced to two disjoint intervals, where the exact modular hamiltonian contains nonlocal terms and can be computed from the results in Refs. \cite{Casini:2009vk,Arias:2018tmw}.

\vspace{10pt}

%\textbf{Acknowledgments:}
%It is a pleasure to thank Clifford V. Johnson, Chris J. Fewster, Nick Warner,  Krzysztof Pilch and Robert Walker for useful comments and discussions. This work was partially funded by the US Department of Energy under grant de-sc 0011687.
\begin{acknowledgments}
It is a pleasure to thank Clifford V. Johnson, Chris J. Fewster, Nicholas P. Warner, Krzysztof Pilch, and Robert Walker for useful comments and discussions. This work was partially supported by the U.S. Department of Energy under Grant No. DE-SC0011687.
\end{acknowledgments}
\appendix

%%%%%%%%%%%%%%%%%%%%%%%%%%%%%%%%
\section{Two-dimensional CFT}
\label{sec:apA}
%%%%%%%%%%%%%%%%%%%%%%%%%%%%%%%%

In this Appendix, we present an independent field theory derivation and generalization of the energy inequality (\ref{eq:38}) for two-dimensional CFTs. To do so, we use the following result,
\begin{equation}\label{eq:5}
\int_{-\infty}^{+\infty}
  dx\,h(x)\langle T_{00}(x) \rangle
  \ge 
  -\frac{c}{6\pi}
  \int_{-\infty}^{+\infty}dx\,
  \left(
  \frac{d}{dx}\sqrt{h(x)}
  \right)^2\ ,
\end{equation}
rigorously derived by Fewster and Hollands for a general CFT \cite{Fewster:2004nj}. The central charge is given by $c$, while $h(x)$ is a non-negative and even\footnote{The inequality can also be written for noneven weight functions; see Ref. \cite{Fewster:2004nj}.} weight function that belongs to the Schwartz space. For a fixed function $h(x)$, the bound on the right-hand side must hold for the expectation value on any state. In Ref. \cite{Fewster:2004nj}, it was shown that the bound is sharp, meaning that for a given function $h(x)$ there is always a state which saturates the inequality. An extension involving mixed states was derived in Ref. \cite{Blanco:2017akw} from the monotonicity property of relative entropy.

We now take the function $h(x)$ equal to the local temperature in Eq. (\ref{eq:37}) inside $A$ and zero outside, so that the left-hand side of (\ref{eq:5}) becomes the expectation value of the modular hamiltonian
\begin{equation}\label{eq:4}
\langle K_A \rangle \ge -
  \frac{c}{6\pi}
  \int_{-R}^{R}dx\,
  \left(
  \frac{d}{dx}\sqrt{f(x)}
  \right)^2\ ,
\end{equation}
where $f(x)=\pi(R^2-x^2)/R$. Since this bound is sharp, calculating the right-hand side will give an expression for the modular vacuum energy $k_0$. Changing variables to $u=x/R$ and using that the integrand, is even we find
\begin{equation}\label{eq:8}
\langle K_A \rangle \ge 
  -\frac{c}{3}\int_0^1du
  \left(\frac{u^2}{1-u^2}\right)\ .
\end{equation}
The resulting integral is infinite due to the contribution when $u\rightarrow 1$, precisely where the chosen function $h(x)$ is nondifferentiable. This is exactly what we expected from our discussion in Sec. \ref{sec:2}; the lower bound on the modular hamiltonian is divergent due to infinite negative energy contributions at the boundary. In order to extract a sensible result, we introduce a regulator $\epsilon$ according to $u_{\rm max}=1-\epsilon/R $, so that the integral can be easily solved,
$$
\langle K_A \rangle \ge
  -\frac{1}{2}
  \left[
  \frac{c}{3}\ln\left(
  \frac{2R}{\epsilon}   
  \right)-\frac{2}{3}c
  \right]
  \ ,
$$
where we have only kept the divergent and finite terms in the $\epsilon/R \rightarrow 0$ limit. Between square brackets, we recognize the entanglement entropy of the ground state reduced to a segment of length $\ell=2R$ \cite{Holzhey:1994we,Calabrese:2004eu}. The constant term is a nonuniversal contribution which can be absorbed into a redefinition of the regulator according to $\epsilon \rightarrow e^2\epsilon$. We then have the following result:
\begin{equation}\label{eq:24}
\langle K_A \rangle \ge k_0=-\frac{1}{2}S(\rho_A)\ .
\end{equation}	
This inequality agrees with the one obtained by calculating the right-hand side of (\ref{eq:38}) using that $S_{\infty}(\rho_A)=S(\rho_A)/2$ from Refs. \cite{Calabrese:2004eu,Cardy:2016fqc}. It also matches with the holographic calculation in (\ref{eq:12}).

This procedure for calculating the modular vacuum energy will be useful whenever the modular hamiltonian is proportional to the energy-momentum tensor. For a global thermal state reduced to an interval of length $\ell=2R$, this is also the case, but with inverse local temperature equal to \cite{Borchers:1998ye,Wong:2013gua,Cardy:2016fqc}
\begin{equation}\label{eq:13}
f_\beta(x)=
  \frac{2\beta \sinh\big(\pi(R-x)/\beta\big)\sinh\big(\pi(R+x)/\beta\big)}
  {\sinh\left(2\pi R/\beta\right)}\ .
\end{equation}
Considering (\ref{eq:5}) with $h(x)=f_\beta(x)$ inside the interval and zero outside, we get the modular hamiltonian on the left-hand side and an integral on the right, which after the change of variables $z=\coth(\pi R/\beta)/\coth(\pi x/\beta)$ is reduced to
$$
\langle K_A^\beta \rangle   \ge -\frac{c}{6}\left[
  \int_{0}^{1}
  \frac{2dz}{1-z^2} -
  \left(\frac{2\pi R}{\beta}\coth\left(\frac{2\pi R}{\beta}\right)
  +1\right)
  \right] \ .
$$
Once again, we obtain a divergent integral due to the nondifferentiability of the function at the boundary. To regulate such divergence, we introduce a regulator which takes into account the change of coordinates, $z_{\rm max}=\coth(\pi R/\beta)/\coth(\pi(R-\epsilon)/\beta)$,
so that the integral can be easily solved and gives
\begin{subequations} 
\begin{eqnarray}
\nonumber
\langle K_A^\beta \rangle  \ge 
  -\frac{1}{2}
  \left[
  \frac{c}{3}
  \ln\left(\frac{\beta}{\pi \epsilon}
  \sinh\left(
  \frac{2\pi R}{\beta}
  \right)\right) 
  -\frac{2}{3}c\right]
  + 
  \\[5pt]\nonumber
\frac{c}{6}
  \left[
  \frac{2\pi R}{\beta}
  \coth\left(\frac{2\pi R}{\beta}\right)-
  1
  \right]
  \ .
\end{eqnarray}
\end{subequations}
The first term between square brackets we recognize as the entanglement entropy of the thermal state reduced to a segment of length $\ell=2R$ \cite{Calabrese:2004eu}, where we identify the same nonuniversal constant factor we had for the ground state.

The second term can be correctly identified as $\langle K_A^\beta \rangle_\beta$ after solving a simple integral and using that the energy density of a thermal state is given by $\langle T_{00}(x)\rangle_\beta=c\pi/6\beta^2$ \cite{DiFrancesco:1997nk}\footnote{This is computed by compactifying the time direction into a circle of radius $\beta/2\pi$ (which maps the vacuum into a thermal state) and using that the energy-momentum tensor transforms according to the Schwartzian derivative.}. We then find the following inequality:
\begin{equation}\label{eq:14}
\langle K_A^{\beta} \rangle\ge k_0(\beta)=-\frac{1}{2}
  S(\rho_A^\beta)
  +\langle K_A^{\beta} \rangle_\beta\ .
\end{equation}
Comparing with the general expression of $k_0$ given in (\ref{eq:36}) and using that $S_{\infty}(\rho_A^\beta)=S(\rho_A^\beta)/2$ from Refs. \cite{Calabrese:2004eu,Cardy:2016fqc}, we find perfect agreement with our previous discussion. 

The divergent contribution to the modular vacuum energy $k_0(\beta)$ in both the zero (\ref{eq:24}) and finite temperature (\ref{eq:14}) cases	 is independent of $\beta$ and therefore exactly the same. Having argued that such divergence has its origin in the nondifferentiability of the weight function at the boundary, we expect $f_\beta(x)$ to be independent of $\beta$ near $x=\pm R$. Taylor expanding (\ref{eq:13}), we find that this is indeed so:
$$
f_\beta(x)=\pm 2\pi(x \pm R)+\mathcal{O}(x\pm R)^2\ .
$$

%%%%%%%%%%%%%%%%%%%%%%%%%%%%%%%%
\section{Black hole thermodynamics}
\label{sec:apB}
%%%%%%%%%%%%%%%%%%%%%%%%%%%%%%%%

In this Appendix, we briefly review the calculation of the zero temperature mass and entropy of the hyperbolic black hole for Einstein gravity in $(d+1)$ space-time dimensions. The black hole solution is given by \cite{Emparan:1999gf}
$$ds^2=-V(r)\left(dt L /R\right)^2+
  dr^2/V(r)+r^2dH_{d-1}^2\ , $$
where $dH_{d-1}^2$ is the unit metric on the $(d-1)$ hyperbolic plane and $L$ is the AdS radius. The time coordinate has been rescaled so that in the limit $r\rightarrow +\infty$ the boundary metric $\mathbb{R}\times \mathbb{H}^{d-1}$ is recovered with curvature scale $R$. 

The function $V(r)=(r/L)^2-1-\mu/r^{d-2}$ determines the horizon radius $r_+$ from $V(r_+)=0$, while the black hole mass is related to the factor $\mu$ according to
$$M_{\rm BH}=\frac{(d-1)w_{d-1}}{2\ell_p^{d-1}}
\frac{L\mu}{R}=
  \frac{(d-1)w_{d-1}Lr_+^{d-2}}{2\ell_p^{d-1}R}
  \left[\left(\frac{r_+}{L}\right)^2-1\right]\ ,$$
where $w_{d-1}$ is the infinite volume of the unit hyperbolic plane, $\ell_p$ is Planck's length, and in the second equality we have written $\mu=\mu(r_+)$ from $V(r_+)=0$. The temperature of the black hole can be computed from the surface gravity $\kappa$ as 
\begin{equation}\label{eq:18}
T=
  \frac{\kappa}{2\pi}=\frac{V'(r_+)L}{4\pi R}=
  \frac{(d-2)}{4\pi  R}\frac{L}{r_+}
  \left[\frac{d}{d-2}\left(\frac{r_+}{L}\right)^2-1\right]\ ,
\end{equation}
where it is equal to $\tilde{T}=1/(2\pi R)$ for $r_+=L$. From the first law of black hole thermodynamics $dS=dM/T$, we can compute its entropy as $S=2\pi w_{d-1}\left(r_+/\ell_{p}\right)^{d-1}$. 

From (\ref{eq:18}), we can solve for the zero temperature horizon radius and find $(r_+^0/L)^2=(d-2)/d$, so that the zero temperature mass and entropy are given by
\begin{subequations} 
\begin{eqnarray}
M_{\rm BH}(0)=&
  \left(\frac{1-d}{d}\right)
  \left(\frac{d-2}{d}\right)^{(d-2)/2}
  \tilde{T} S_{\rm BH}(\tilde{T}) \ ,
  \nonumber \\ [5pt]
S_{\rm BH}(0)=&
  \left(\frac{d-2}{d}\right)^{(d-1)/2}S_{\rm BH}(\tilde{T})\ , \nonumber
\end{eqnarray}
\end{subequations}
where we have written everything in terms  $S_{\rm BH}(\tilde{T})$. From (\ref{eq:27}), we see that the black hole entropy at $\tilde{T}$ is mapped to the entanglement entropy $S(\rho_A)$ (after proper regularization of $w_{d-1}$ \cite{Casini:2011kv}), so that we recover (\ref{eq:20}).

For the hyperbolic black hole in Gauss-Bonnet gravity \cite{Lovelock:1971yv,Cai:2001dz,Nojiri:2001aj}, the calculation is completely analogous but more involved. Following a similar procedure as in the Einstein case (and using the convenient conventions of Ref. \cite{Hung:2011nu}), both the zero temperature mass and entropy can be computed analytically for arbitrary $d$.

For $d=4$, the allowed range of $n_c$ is usually taken as $n_c\in[2/3,2]$ \cite{Hung:2011nu}. However, this does not take into account the fact that any physical black hole solution must have non-negative entropy. With this under consideration, we find $n_c\in[2/3,1+\sqrt{2/3}]$ where for $n_c^{\rm max}$ (\ref{eq:23}) vanishes.

\bibliography{Holography_of_negative_energy_states}
\end{document}